\documentclass[titlepage]{article}

\usepackage[english]{babel}
\usepackage[table,xcdraw]{xcolor}
\usepackage{array}
\usepackage{caption}
\usepackage{sidecap}
\usepackage{csquotes}
\usepackage{svg}
\usepackage{authblk}

\usepackage[
backend=biber,
style=nature,
]{biblatex}
\usepackage[a4paper,margin=1in]{geometry}

\usepackage{amsmath}
\usepackage{siunitx}
\usepackage{graphicx}
\usepackage{subcaption}
\usepackage[colorlinks=true, allcolors=blue]{hyperref}
\title{Diffusion Tensor MRI and Spherical-Deconvolution-Based Tractography on an Ultra-Low Field Portable MRI System}

\author[1]{James Gholam}
\author[1]{Phil Schmid}
\author[1]{Joshua Ametepe}
\author[1]{Alix Plumley}
\author[2]{Leandro Beltrachini}
\author[3]{Francesco Padormo}
\author[3]{Rui Teixeira}
\author[3]{Rafael O'Halloran}
\author[4]{Kaloian Petkov}
\author[4]{Klaus Engel}
\author[5]{Steven CR Williams}
\author[6]{Sean Deoni}
\author[1]{Mara Cercignani}
\author[1]{Derek K Jones}
\affil[1]{Cardiff University Brain Research Imaging Center, Cardiff University, Cardiff, UK}
\affil[2]{School of Physics and Astronomy, Cardiff University, Cardiff, UK}
\affil[3]{Hyperfine, Inc., Guilford, CT, United States}
\affil[4]{Cinematic Rendering, Siemens Healthineers AG}
\affil[5]{Centre for Neuroimaging Sciences, King’s College London, London, UK}
\affil[6]{Bill \& Melinda Gates Foundation, MNCH D\&T, Seattle, WA, USA}
\date{}
\begin{document}
\maketitle

\begin{abstract}
Ultra-low-field (ULF) MRI is emerging as an alternative modality to high-field (HF) MRI due to its lower cost, minimal siting requirements, portability, and enhanced accessibility—factors that enable large-scale deployment. Although ULF-MRI exhibits lower signal-to-noise ratio (SNR), advanced imaging and data-driven denoising methods enabled by high-performance computing have made contrasts like diffusion-weighted imaging (DWI) feasible at ULF. This study investigates the potential and limitations of ULF tractography, using data acquired on a 0.064 T commercially available mobile point-of-care MRI scanner. The results demonstrate that most major white matter bundles can be successfully retrieved in healthy adult brains within clinically tolerable scan times. This study also examines the recovery of diffusion tensor imaging (DTI)-derived scalar maps, including fractional anisotropy and mean diffusivity. Strong correspondence is observed between scalar maps obtained with ULF-MRI and those acquired at high field strengths. Furthermore, fibre orientation distribution functions reconstructed from ULF data show good agreement with high-field references, supporting the feasibility of using ULF-MRI for reliable tractography. These findings open new opportunities to use ULF-MRI in studies of brain health, development, and disease progression—particularly in populations traditionally underserved due to geographic or economic constraints. The results show that robust assessments of white matter microstructure can be achieved with ULF-MRI, effectively democratising microstructural MRI and extending advanced imaging capabilities to a broader range of research and clinical settings where resources are typically limited.
\end{abstract}
\newpage
\section{Introduction}

Diffusion MRI and tractography are widely used in neuroscience for developmental studies, tracking disease progression, and examining tract spatial properties across various populations~\cite{DellAcqua2024}. This broad interest in microstructural assessment and virtual dissection~\cite{Catani2002} has driven significant advances in MRI methods at high field, and in this work we seek to demonstrate the useful application of some of these techniques at ultra-low fields. 

 In contrast to high field (\qty{>1}{T}) MR systems, ULF systems (\qty{<0.1}{T}) are free from numerous constraints imposed by superconducting systems, including non-reliance on cryogenic cooling or high voltage power supplies, and tolerance of power supply interruptions. Alongside significant reductions in weight, $B_0$ fringe fields, and RF heating, ULF systems often have exceptional mobility. They can be easily moved between, for example, ICU settings, remote locations, or directly to emergency sites—and in many cases, they only need to be plugged into the local mains circuit or a battery supply to operate. As well, the reduced size, power and siting requirements, and cost of ULF-systems make them attractive in research contexts, where a high-field system may represent an unacceptable cost burden to a research endeavour.
\sisetup{range-units = single, range-phrase = --}
The majority of contemporary diffusion-based tractography studies are conducted using data acquired on high-field MRI systems (e.g. \qty{3}{\tesla}) with moderate to high gradient amplitudes \qtyrange{40}{80}{\milli \tesla \per \meter}), enabling rapid diffusion weighted single-shot (spin) echo planar imaging (ssDW-EPI)~\cite{Tax2022}. While ssDW-EPI is robust and insensitive to motion within each shot, numerous factors complicate its use at ULF. ULF DTI was first demonstrated employing non-EPI readouts~\cite{Plumley2022}, and subsequently using DW-EPI~\cite{Ding2024}. The performance of fast EPI readouts are limited by low gradient strength and high $B_0$ inhomogeneity in ULF systems, causing significant distortions at higher resolutions, and significant signal dephasing arising from low bandwidth in phase encoding directions. Long echo trains combined with slower and weaker gradients require long echo times. This produces significant $T_2$-related signal loss, and while this may be mitgated by high readout bandwidth, this further degrades SNR. Additionally, at ULF, Johnson noise in inductive detector coils is the dominant noise component over body noise. This results in 3D imaging offering higher SNR at the point of detection due to the larger volume excited over 2D imaging. In contrast, the minimal SAR produced by short RF pulses at ULF allow 3D fast, multi-echo sequences to be run without exceeding SAR limits, and mitigate field inhomogeneity induced dephasing and distortion while maximising SNR.

 The principal challenge hindering widespread use of ULF systems for DWI is the SNR available per unit time. As the contrast mechanism of DWI relies on signal attenuation~\cite{Jones2010}, obtaining acceptable data at the SNRs typical of ULF systems is particularly challenging. Though clinically viable ULF diffusion imaging protocols for diagnosis of cerebral ischemia and infarction~\cite{Cahn2020, Cahn2024} have recently become available, extending this to tractography is challenging as it requires high accuracy, which diagnostic DWI may not, in order to be clinically useful. Diffusion-based tractography relies upon unbiased measures with sufficient resolution to discriminate between regions containing white matter and those containing grey matter or CSF. Noise biases diffusion measurement, with Rician noise resulting in reduced apparent fractional anisotropy and mean diffusivity in DTI~\cite{Jones2004}, or, in spherical deconvolution, offsets and increased variance in volume fraction measures and relative fibre angles in mixed fibre populations~\cite{Canales-Rodrguez2015}. Non-uniformity of gradient encoding and/or the static $B_0$-field also complicates the unbiased interpretation of the diffusion signal. 
 
 While advances in signal detection, filtration, digitisation, reconstruction and downstream processing have enabled accelerated single-volume ULF diffusion acquisitions, significant technical obstacles remain to the implementation of a robust tractography protocol at ULF. In this work, we address some of these obstacles, and demonstrate that tractography may be performed with data acquired in feasible scan times at a magnetic field strength of \SI{64}{\milli \tesla} on a commercial ULF-MRI system.

\section{Methodology}
\subsection{Participants}
Five healthy adult volunteers were recruited for this study, with ethical approval from the Cardiff University School of Psychology ethics committee. We targeted a protocol length of 1 hour (standard protocol length for research studies in our centre), additionally aiming to prevent overheating of the passively cooled gradient hardware, which is particularly stressed by the demands of diffusion-encoding pulsed gradients.

\subsection{Data Acquisition}

The scanner employed was a \SI{64}{\milli \tesla} Swoop (hardware version 1.7, software version rc9.0 beta 1) permanent-magnet ULF-MRI system from Hyperfine Inc., equipped with a 3-axis gradient set with peak gradient amplitudes of  X: \SI{24.9}{\milli \tesla \per \meter}, Y: \SI{24.4}{\milli \tesla \per \meter} and  Z: \SI{25.7}{\milli \tesla \per \meter}, at slew rates of \qtylist[list-units = single]{23; 22; 67}{\tesla \per \meter \per \second} respectively. Prior to imaging, the scanner was $B_1$ power calibrated using the manufacturer's standard pre-scan calibration, and $B_0$ shimmed, using linear (gradient) $B_0$ shimming, and measured to have a $B_0$ homogeneity of \SI{1100}{ppm} in a 16cm diameter spherical volume. The single-yoke permanent magnet array is asymmetric in the RL-direction, producing a spatially non-uniform $B_0$ field.

The DWI sequence employed for this study was a 3D multi-shot diffusion weighted, non-Carr-Purcell-McGill-Bloom (CPMG), self-navigated fast-spin-echo, employing a non-Cartesian k-space trajectory, with centre-out phase encoding ordering~\cite{OHalloran_2022}. 212 shots were used for each volume, using a split echo train with 70 echoes across a TR of \SI{800}{\milli \second}, spanning 35 unique phase encodings per TR. The effective echo time was \SI{84}{\milli \second}. The field of view (FOV) was \SI[parse-numbers=false]{220 \times 200 \times 180}{\milli \meter}, in AP/SI/RL directions respectively. Data were reconstructed at a resolution of \SI[parse-numbers=false]{3 \times 3 \times 3}{\milli \meter}. The readout was fully sampled and oriented in the RL direction, whereas the phase encoding directions were oversampled by \SI{166}{\percent}.Oversampling ratios from $400\%$ to $100\%$ (Nyquist limit) were tested. Higher oversampling ratios were found to increase sensitivity to subject motion, manifesting as diffuse blurring, and linearly increased scan time. Minimal oversampling resulted in regions of uniform signal and artificially sharp edges that did not correspond to anatomical tissue boundaries.

Diffusion sensitisation utilised a monopolar pulsed-gradient spin-echo scheme, with b=\SI[per-mode=symbol]{945}{\second \per \milli \meter \squared} at isocentre, with diffusion encoding gradients of duration $\delta$ = \SI{35}{\milli \second} and separation $\Delta$ = \SI{42}{\milli \second}. The non-diffusion weighted sequence  employed $2\times$ phase oversampling to provide a high SNR for apparent diffusion coefficient (ADC) calculation, and lasted 94 seconds. Each diffusion-weighted volume (DWI), to collect all 212 shots, required 170 seconds of acquisition. Eighteen diffusion encoded volumes were acquired with isotropically distributed axes organised into three groups of six. The encoding axes in each group of six were arranged using an electrostatic repulsion algorithm~\cite{Jones2004}, enabling early scan termination if necessary while still permitting a estimation of the full diffusion tensor.

\begin{figure}
    \centering
    \includegraphics[width=1.0\linewidth]{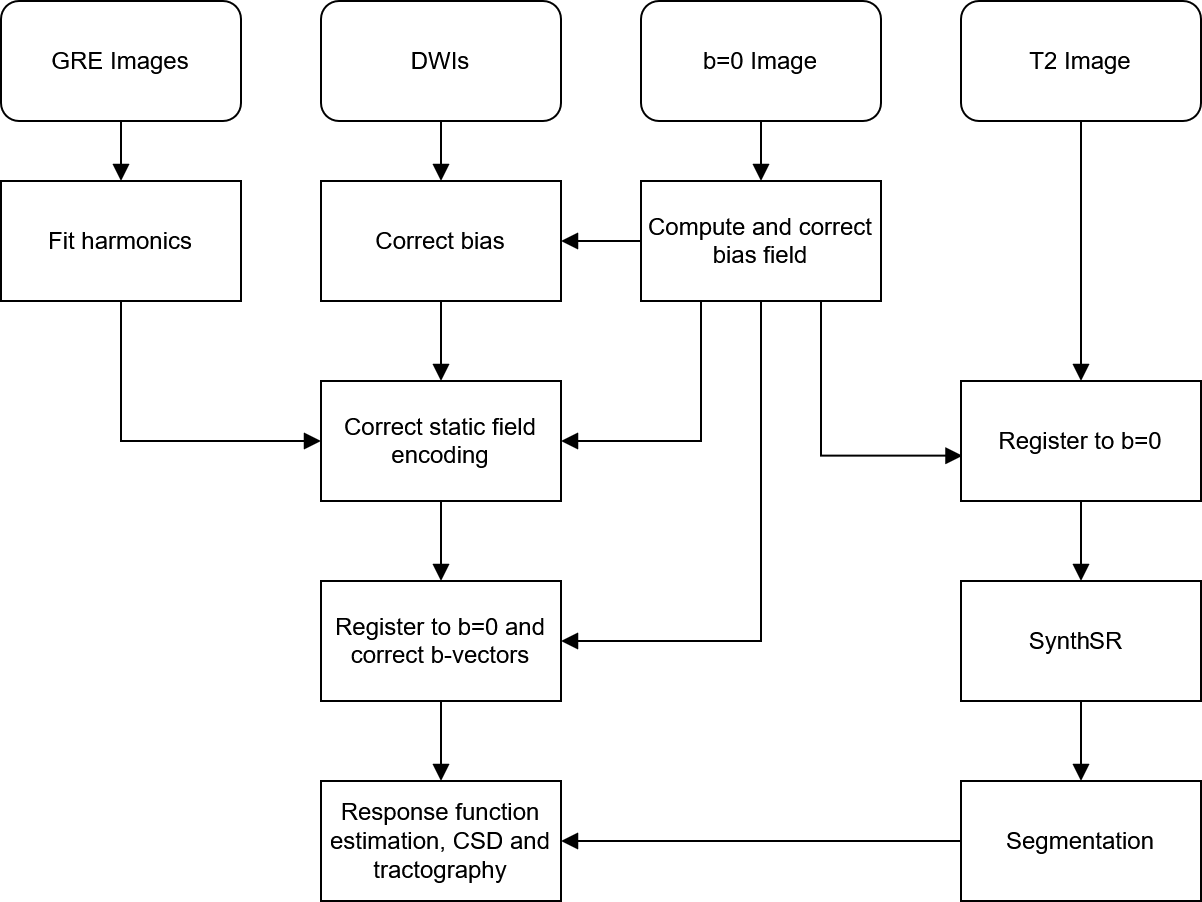}
    \caption{Schematic description of the processing pipeline employed to generate corrected and combined anatomical and DW images used for subsequent tractography}
    \label{fig:pipeline}
\end{figure}

Within the same protocol, a $T_2$-weighted split-echo fast-spin echo sequence was collected at a resolution of \SI[parse-numbers=false]{2 \times 2 \times 2}{\milli \meter} across the same FOV to provide complementary structural information to diffusion measures. The TE was \SI{180}{\milli \second}, the TR was \SI{1600}{\milli \second}, readout bandwith \SI{120}{\kilo \hertz}, and echo train length 64. The data were reconstructed, denoised, debiased and distortion corrected, using the manufacturer methods. The sequence duration was 9 minutes 15 seconds, and these data (which placed less demands on the gradient hardware than DWI) were collected amongst the DW volumes to provide an opportunity for the gradient hardware to cool. 

Between every set of 2 DWIs, the scanner centre frequency $f_0$ was recentered following a short calibration scan to account for $B_0$ drift induced by gradient heating. Data were first filtered and corrected with the vendor's phase self-navigation, online shot rejection and resampling, and their proprietary deep learning reconstruction was used for image formation. Data were distortion corrected for both static field inhomogeneity and gradient induced spatial encoding errors using the manufacturer's pre-computed image-space unwarping method. The built-in bias correction was disabled to facilitate estimation of the bias field using the b=\SI[per-mode=symbol]{0}{\second \per \milli \meter \squared} image. The diffusion images were denoised by the vendor deep-learning algorithm separately to reconstruction.

As a result of the highly inhomogeneous $B_0$ field, in the scanner employed, a large permanent magnetic field gradient was present in FOV. Modelling of this gradient predicted a significant contribution to the diffusion encoding, which if characterised could be retrospectively corrected for. A prospective fieldmap was obtained prior to in vivo scanning using a gradient recalled echo image of a flood fill phantom, measured at 2 different echo times to characterise the major $B_0$ field variations.

\subsection{Data Analysis}
\subsubsection{\texorpdfstring{$\mathbf{B_1}$}{B1}-bias field correction}
Reconstructed DWIs were first corrected for $B_1$ bias by estimation of the field on the b=\SI[per-mode=symbol]{0}{\second \per \milli \meter \squared} image using the N4 bias field correction algorithm~\cite{Tustison2010}, and the inverse of this field applied to the b=\SI[per-mode=symbol]{0}{\second \per \milli \meter \squared} and all DWIs in the series~\cite{Tax2022}. 

\subsubsection{Correction for Background Field Gradient (\texorpdfstring{$\mathbf{B_0}$}{B0} Inhomogeneity)}
Images were retrospectively corrected to account for the large permanent magnetic field gradient, using the b=\SI[per-mode=symbol]{0}{\second \per \milli \meter \squared} image as a reference to separate encoding from other contrast effects (see Appendix). In the scanner employed in the current work, at positions \SI{8}{\centi \meter} from the isocentre, this gradient may be up to \SI{1.4}{\milli \tesla \per \meter}, i.e. up to 7\% of the specified diffusion encoding gradient. this can lead to errors of 16.1\% in the ADC. The contribution of this field inhomogeneity may be modelled as an additive magnetic field gradient producing a spatially dependant linear scaling to the prescribed b-value by a factor $a(\mathbf{r})$. The diffusion-weighted signal $S'$ from a voxel with apparent diffusion coefficient (ADC) $D$, with the contribution from the spurious field may be modelled as
\begin{equation}
    S'(\mathbf{r}) = S_{0}(\mathbf{r})\exp({-a(\mathbf{r})bD(\mathbf{r})})
\end{equation}
with a correction depending on $a$ and the signal from the b=\SI[per-mode=symbol]{0}{\second \per \milli \meter \squared} image $S_0$. The correction may be represented as a simple multiplication:
\begin{equation}\label{eq:diff_correction_body}
     S(\mathbf{r})= \exp\left(\log\left(\frac{S'(\mathbf{r})}{S_0(\mathbf{r})}\right) \left(1 - \frac{1}{a(\mathbf{r})}\right)\right),
\end{equation}
where $S(\mathbf{r})$ is the corrected signal at a given position.

\subsubsection{Registration and Superresolution}
The diffusion-weighted volumes were affinely registered to the non diffusion-weighted volume using the hierarchical ANTs implementation using a mutual information cost function, and the rotation component of the affine transform applied to the respective volumes b-vectors~\cite{Leemans2009}. The structural $T_2$ weighted scan was registered to the b=\SI[per-mode=symbol]{0}{\second \per \milli \meter \squared} image, then super-resolved (SR) to a $T_1$ weighted contrast using SynthSR~\cite{Iglesias2021, Iglesias2022, Iglesias2023} to \SI[parse-numbers=false]{1 \times 1 \times 1}{\milli \meter} resolution. This structural image used as the basis for masking, partial volume calculations and in anatomically constrained tractography~\cite{Smith2012}.

\subsubsection{Tissue Segmentation}
The SR $T_1$-weighted image was segmented using Freesurfer's recon-all pipeline~\cite{Collins1994, dale:99, fischl:99, FischlLiuDale, FischlSalat2002, Fischl2004S69} to give masks of all major structures in the brain in the same space as the diffusion data. The high resolution information afforded by this approach is desirable for the creation of grey- and white-matter masks that allow inference of the partial volumes present in the lower resolution diffusion data. A 5 tissue-type (5tt) segmentation was constructed from this, as was a grey-matter white-matter interface (GMWMI) mask. Manual voxel selection in the corpus callosum midline was used to estimate single-fibre response functions with spherical harmonic order 4 (to match the number of equations solved by constrained spherical deconvolution (CSD) to the number of DWIs) using mrtrix3~\cite{Tournier2019}.

\subsubsection{DTI, Fibre Orientation and Tractography}
Diffusion tensors were fitted to the observed data using the RESTORE method~\cite{Chang2005, Chung2006, Yendiki2014} implemented in DIPY Version 1.11~\cite{Garyfallidis2014}. Noise levels were initially estimated with the DIPY "estimate\_sigma" method~\cite{Koay2006, Park2009}, then manually adjusted to maximise visual consistency.

The response functions permitted estimation of fibre-orientation distributions for each tissue type using the MSMT-CSD~\cite{Jeurissen2014} method given in mrtrix3. This was then used alongside the 5tt masks and the GMWMI masks to perform anatomically constrained tractography using the mrtrix3 implementation of the iFoD2 algorithm. Final tractograms were computed using three methods:
\begin{enumerate}
\item Manual definition of inclusion and exclusion ROIs for major WM bundles, using the super-resolved $T_1$ to estimate anatomically informed regions of interest~\cite{Catani2002, Conturo1999}. These ROIs were used to filter a 2 million streamline wholebrain iFoD2-ACT tractogram, only including fibres within inclusion ROIs, and trimming fibres within exclusion ROIs.

\subsubsection{Visualisation}
The ultra-low-field MRI data was visualised using Cinematic Rendering~\cite{Comaniciu2016, Dappa2016}, a Monte-Carlo path tracing engine developed by Siemens Healthineers, which integrates various data sources from medical imaging to generate photorealistic images and animations. Fibre data was processed with 3D Slicer~\cite{Fedorov2012} and MRtrix3~\cite{Tournier2019}, transforming MRtrix .tck files into VTK~\cite{Schroeder2006} polygonal data. Fibre probability density maps were generated using MRtrix3's tckmap command, producing directionally encoded colour space fibre data and fibre density volume data. Polygonal fibre data was integrated with T2 volume data using a unified path tracing approach, constructing a bounding volume hierarchy for fast intersection of rays with fibre representations. The Cinematic Rendering engine registers T2-weighted MRI sequences with volumetric fibre representations, applying a brain mask signed distance field to reveal anatomical structures. Users can adjust fibre data density and colour through interactive transfer functions. During path tracing, samples from all volumes and meshes are composited in the shader, using high-dynamic range lighting and tone-mapping for final output. The polygonal fibre method generates many primitives, limiting visualisation to specific bundles, while the volumetric method allows visualisation of the entire fibre set, offering sharper representation and comprehensive coverage. Both methods can be combined for comprehensive coverage while highlighting selected fibre bundles with a precise polygonal representation.

\item  Filtering of a 10-million streamline iFoD2-ACT tracked brain using TractSeg ROIs. Bundle ending segmentations (i.e. spatial estimates of the cortical terminations of WM bundles) were generated using the tool TractSeg~\cite{Wasserthal2018a}, using previously computed fODF peaks in WM. Only tracts which started and ended in bundle endings, and remaining within bundle ROIs were retained, using the corresponding ROIs generated using TractSeg. 

\item Tract-orientation mapping (TOM)~\cite{Wasserthal2018b} using TractSeg derived ROIs and fODF peaks in WM.
\end{enumerate}

\subsubsection{High-field Validation Dataset}
High field measurements were also collected to serve as a high SNR reference standard against ULF measurements. Data were collected on the same participants as for ULF at \SI{3}{\tesla} on a Siemens Connectom MR system with \SI{300}{\milli \tesla \per \meter} gradients using a multiband DW-echo-planar imaging sequence, following a high-angular-resolution diffusion imaging (HARDI) protocol with 253 directions at  b-values of \qtylist[list-units = single]{0;200;500;1200;2400;4000;6000}{\second \per \milli \meter \squared} with shells containing \numlist{13;20;20;30;61;61;61} isotropically distributed directions respectively. Imaging resolution was  \SI[parse-numbers=false]{2 \times 2 \times 2}{\milli \meter}, TR \SI{3000}{\milli \second}, TE \SI{59}{\milli \second}, FOV \SI[parse-numbers=false]{220 \times 220 \times 132}{\milli \meter \cubed}. Diffusion encoding used gradients of duration $\delta$ = \SI{7}{\milli \second} and separation $\Delta$ = \SI{24}{\milli \second}.  The total readout duration was \SI{29}{\milli \second}. Data were corrected for drift using an in house method~\cite{Vos2017}, and similarly for gradient nonuniformity induced distortions. Data were further corrected for susceptibility and eddy current induced distortions using a reverse-phase encoding method~\cite{Andersson2003,Andersson2016a,Andersson2016b, Smith2004}, and denoised with MP-PCA~\cite{Veraart2016}. Gibbs deringing using a subvoxel shift method was additionally employed~\cite{Kellner2016}.  A $T_1$ weighted MP-RAGE structural scan was acquired and processed with Freesurfer's recon-all and mrtrix3 to produce a 5tt segmentation and corresponding grey-matter white-matter interface masks, and these were affinely coregistered to the diffusion data.

The diffusion data were used to compute fODFs, mirroring the methods employed at ULF, though with maximum spherical harmonic l=8, and the Dhollander algorithm~\cite{Dhollander2016} was used to estimate the tissue response functions. Lastly, fODFs were also calculated using MSMT-CSD. As well, 2 b=\SI[per-mode=symbol]{0}{\second \per \milli \meter \squared} and 30 b=\SI[per-mode=symbol]{1200}{\second \per \milli \meter \squared} volumes were extracted from the data and fitted to spherical harmonic lmax=4, identically to the ULF data. This high SNR but low direction count dataset was used to inform on the impact of SNR on fODF measures vs the ULF data. The high direction count and high b-value dataset was used as a "silver standard" measurement.

\newpage
\section{Results}
\subsection{Diffusion and Fibre Orientation Measures}
\begin{SCfigure}
    \centering
    \includegraphics[width=0.70\linewidth]{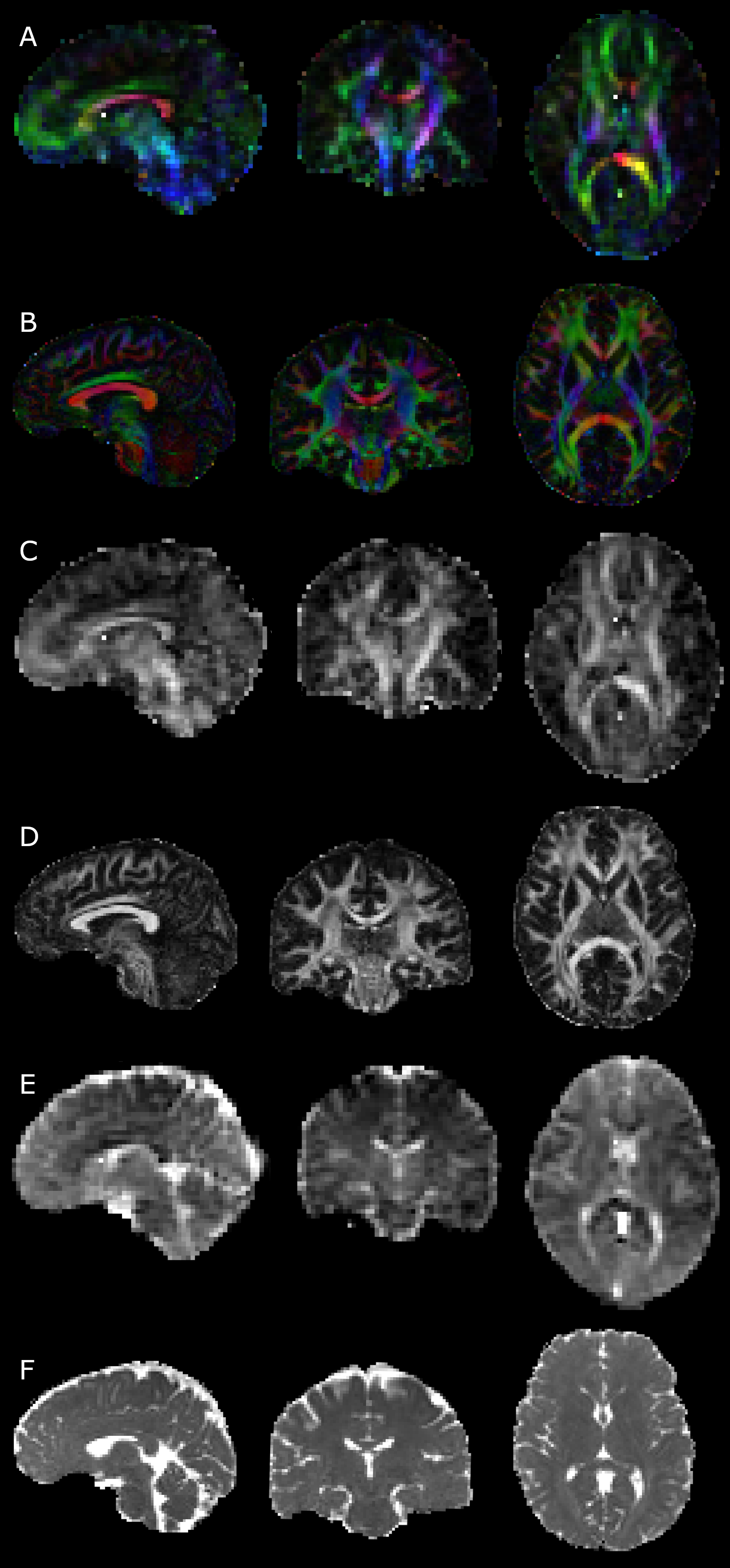}    
    \caption{Comparison of quantitative diffusion between high- (B,D, and F) and ultra-low-field (A, C, and E) measures obtained in the same subject in sagittal, coronal and axial planes. High field measures used 30 directions at b=\SI[per-mode=symbol]{1200}{\second \per \milli \meter \squared}. A,B: diffusion encoded colour maps, weighted by FA. Distinct asymmetry is observed between left and right hemispheres in ULF data. This may arise from different spatial noise dependence, or from uncorrected diffusion encoding nonuniformity. C,D: fractional anisotropy - FA is noticeably lower in ULF measurements, again likely due to the elevated noise floor. Transverse WM is distinctly darker in FA measurements ULF than in HF, possibly due to the broad PSF and partial volume effects causing thinner bundles to appear more diffuse. E,F: mean diffusivity - shading is observed in ULF measurements, but otherwise there is good correspondence between HF and ULF measures.}
    \label{fig:hf_lf_quant}
\end{SCfigure}

 Diffusion measures at ULF visually corresponded with HF measures taken with small numbers of directions and a single shell. Diffusion encoded colour maps (Figure \ref{fig:hf_lf_quant}A, B) showed good agreement in major tracts, however anisotropic tissue further from the mid-sagittal axis plane (and hence isocentre) was not fully recovered, likely due to the high noise floor suppressing apparent fractional anisotropy. This was more visible in ULF FA maps directly (Figure \ref{fig:hf_lf_quant}C, D), where subcortical fibres were generally not observed, though were highly visible at HF. Imperfection and inconsistency in the shape of some major fibre bundles was observed, for example in the major forceps, where one side was measured to be more anisotropic than the other. Mean diffusivity maps (Figure \ref{fig:hf_lf_quant}E, F) showed generally good correspondence. Image quality and subsequent DTI measures and tracking quality was not found to vary substantially between individuals, and the results shown herein are a representative result of a compliant subject. 

\begin{figure}
    \centering
    \includegraphics[width=1.0\linewidth]{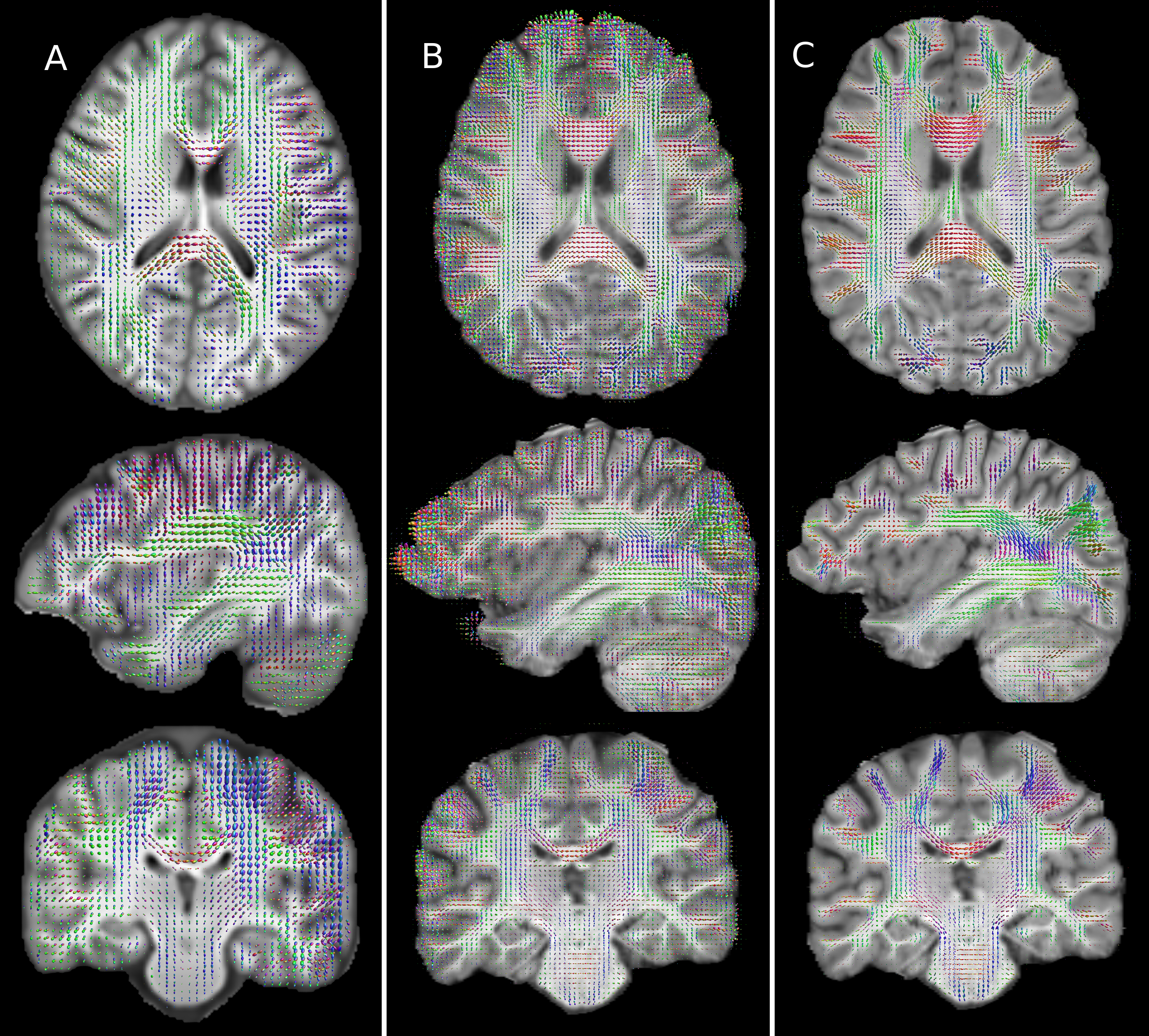}
    \caption{Comparison of fibre orientation distribution plots computed in the same representative subject for the ULF (A), HF with 30 directions and one shell (B), and HF with 253 diffusion volumes across 6 shells (C). Substantial blurring is apparent in A, with margins of WM structures being less well defined and adjacent structures averaging together. Conspicuously this leads to large assigned fibre populations in GM, particularly in highly convoluted regions. Comparing A with C shows there is generally good correspondence between ULF and the gold standard measurement.  Comparing B and C shows the significant impact of a reduced direction count on an otherwise high SNR measurement, with B showing smaller bundles having erroneously large lobes, and grey matter with significant anisotropy that is not corroborated by C.}
    \label{fig:hf_lf}
\end{figure}

Fibre orientation measures were also broadly comparable between HF and ULF (Figure \ref{fig:hf_lf}), though limitations in effective resolution were marked, with large fODF lobes spuriously attributed to grey matter and CSF. Reducing the number of directions and shells available to the HF measurement (Figure \ref{fig:hf_lf}B) demonstrated that the quality of fODF estimation is highly dependent on large numbers of diffusion directions and multiple shells. Despite these limitations, CSD correctly estimated the orientation of some subcortical fibres, even though DTI metrics failed to assign anisotropy to these regions.

\subsection{Tractography}
\begin{figure}
    \centering
    \includegraphics[width=1.0\linewidth]{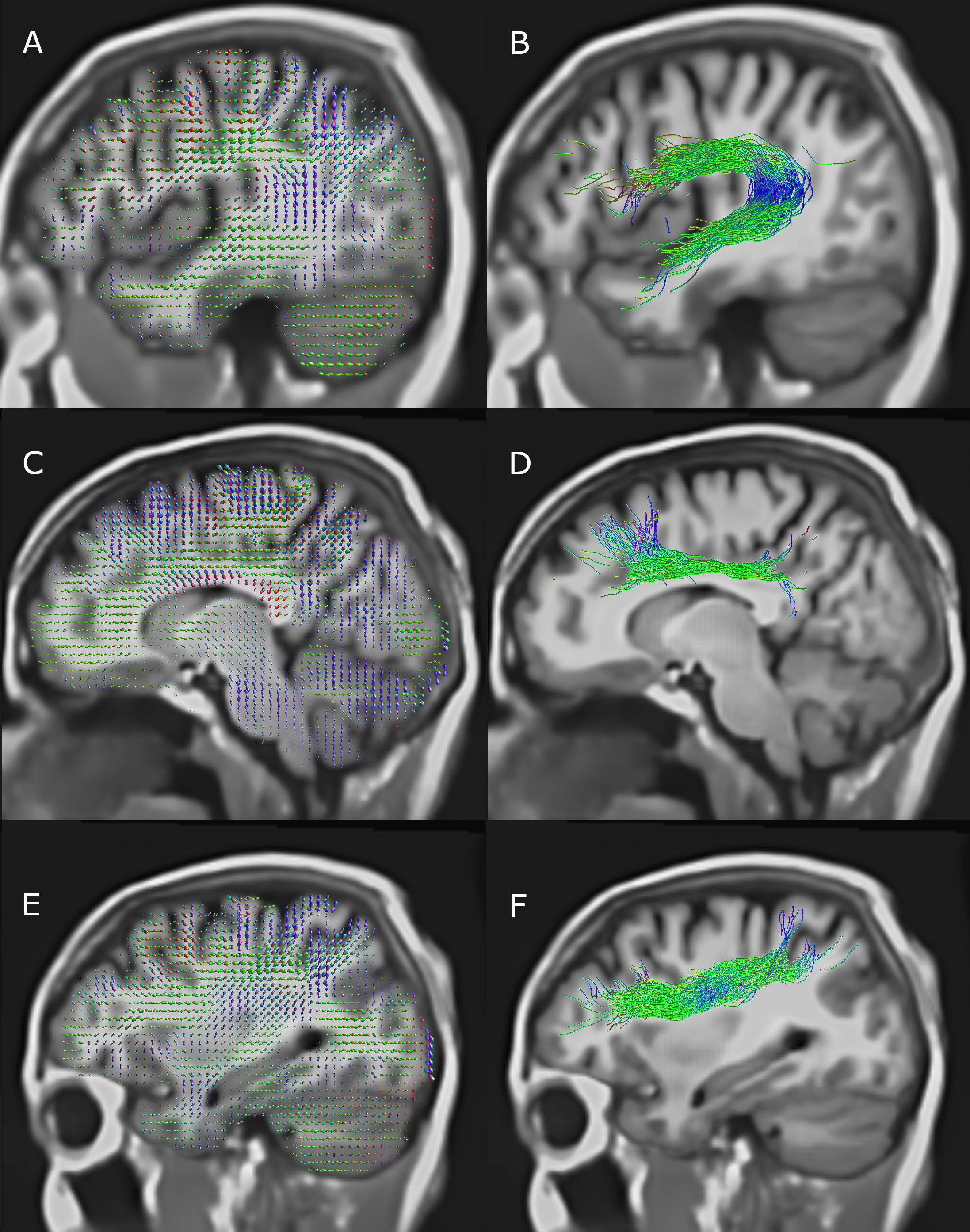}
    \caption{Fibre ODFs and major association tracts retrieved at ULF using manual ROI selection; (A,B) arcuate fasciculus; (C,D) upper portion of cingulum; (E,F) SLF I and II. Tracts shown are not cropped to the displayed slice.}
    \label{fig:association_tracts}
\end{figure}

Tractography derived from orientation maps using manual ROI selection showed good coherence in the core of the fasciculus (where anisotropy is high), but rapid dispersion when moving into more cortical regions(as in Figure \ref{fig:association_tracts}D). Gradual curvature within tracts such the superior longitudinal fasciculus, arcuate fasciculus and corpus callosum (Figures \ref{fig:association_tracts}B and \ref{fig:association_tracts}F) was correctly recovered, suggesting that diffusion data has sufficient SNR to produce stable orientation estimates. 

A significant fraction of major WM tracts were reconstructed, including the arcuate, inferior longitudinal, inferior fronto-occipital, uncinate and superior longitudinal fasciculi. As well, as the cingulum, the upper portion of the fornix, and the pontine and corticospinal tracts were well reconstructed. The major interhemispherical corpus callosum was generally well recovered, but showed pronounced morphological differences to expected anatomy in the proximity of the ventricles where motion artifacts are expected to be more severe.   

\begin{figure}
    \centering
    \includegraphics[width=1.0\linewidth]{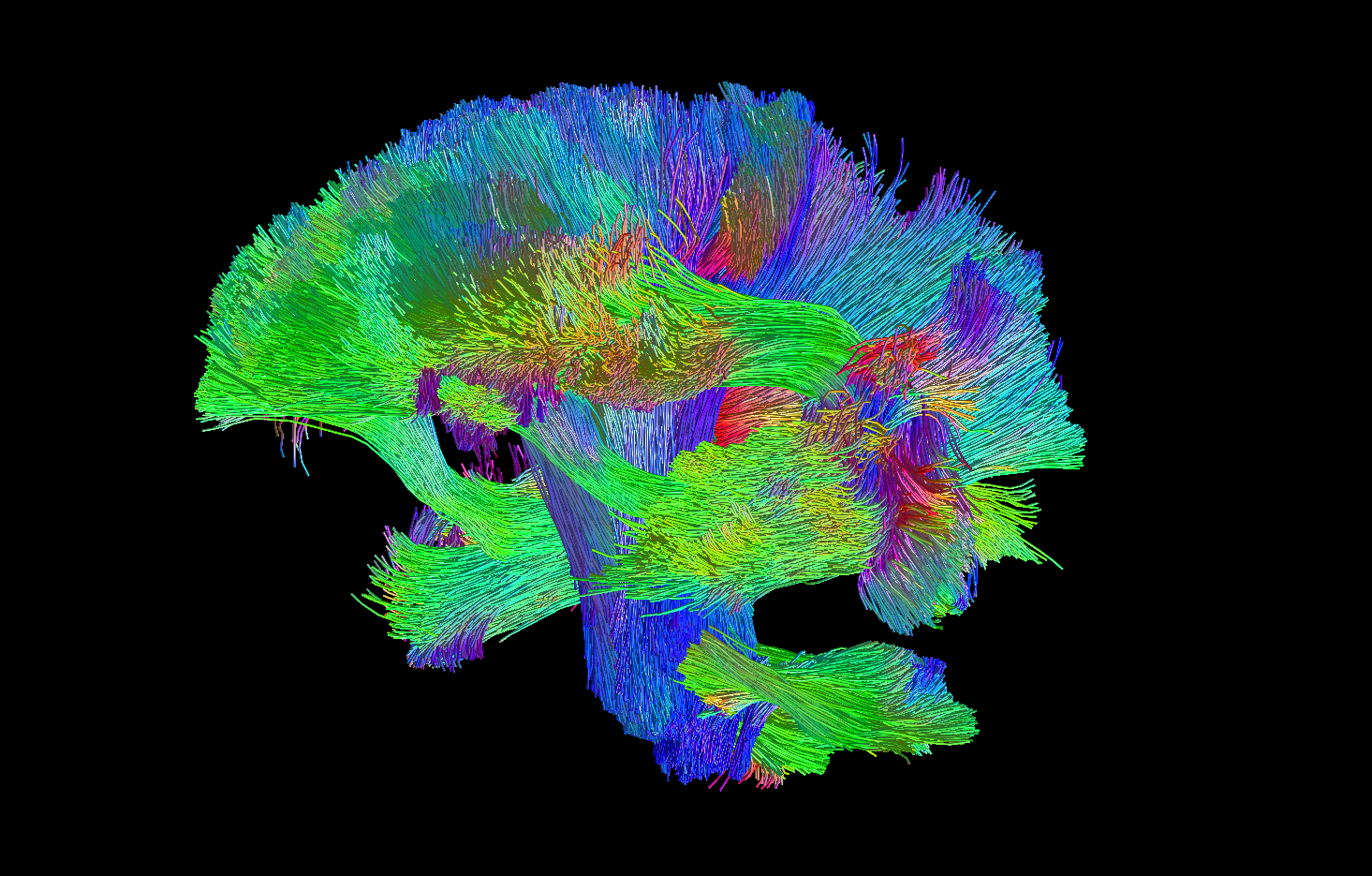}
    \caption{Whole-brain automated tracking conducted with TractSeg.}
    \label{fig:tractseg}
\end{figure}

\begin{figure}
    \centering
    \includegraphics[width=1.0\linewidth]{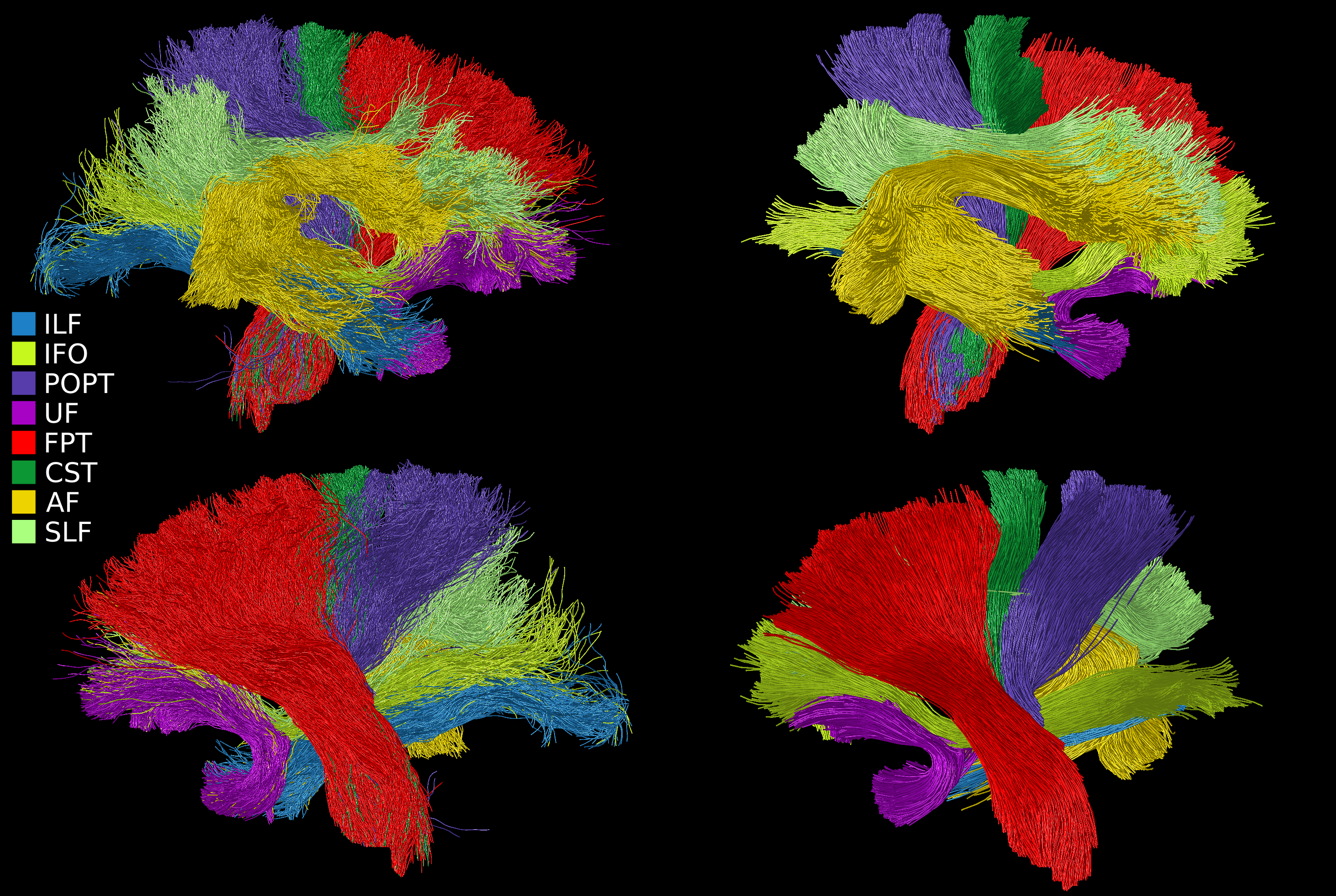}
    \caption{Selection of major WM tracts in a single hemisphere of the brain retrieved using automated ROI filtering of the wholebrain tractogram generated with iFoD2 and ACT (left column) and TOM tracking with TractSeg (right column). Visible tracts are the inferior longitudinal fasciculus (ILF), inferior occipito-frontal fasciculus (IFO), parieto‐occipital pontine (POPT), uncinate fasciculus (UF), fronto-pontine tract (FPT), corticospinal tract (CST), arcuate fasciculus (AF), and superior longitudinal fasciculus (SLF). TOM and automated filtering show close correspondence, though TOM measures are smoother, and truncated upon reaching the cortex.}
    \label{fig:labelled_tractography}
\end{figure}

\begin{figure}
    \centering
    \includegraphics[width=0.8\linewidth]{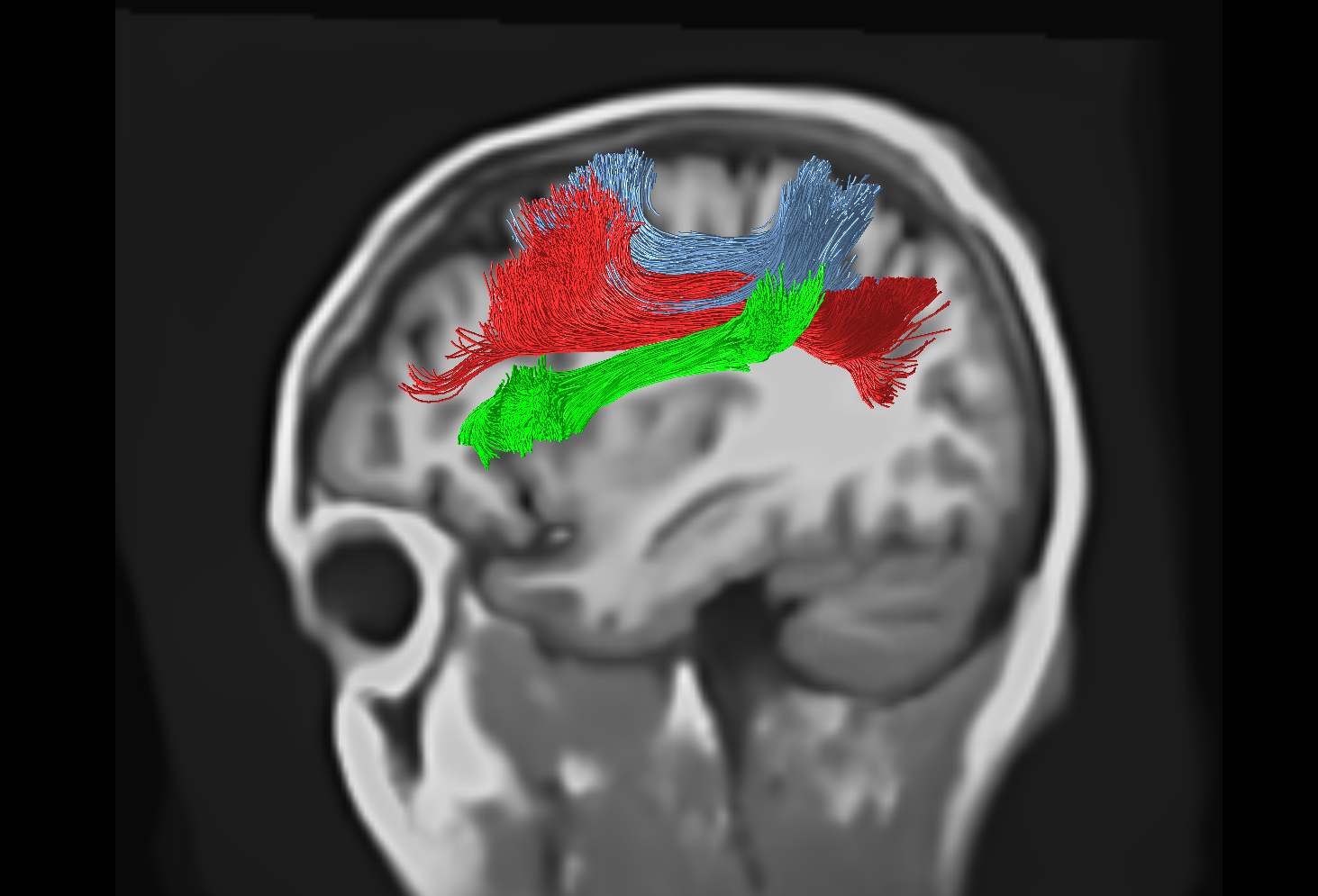}
    \caption{TractSeg-derived tracking of the superior longitudinal fasciculus, reconstructing the SLF I (blue), II (red) and III (green) bundles.}
    \label{fig:slf_tractseg}
\end{figure}

Application of deep-learning priors through TractSeg resulted in much smoother fibre trajectory reconstruction that conformed closer to prior expectations of white matter anatomy (as in Figure \ref{fig:tractseg}), and closely mirrored results obtained using the semi-automated approach using the same ROIs. Tracts obtained using TOM were highly coherent, but the major trunks of bundles still followed approximately the same courses as for CSD-ACT (see Figure \ref{fig:labelled_tractography}). Tracts such as the SLF were reconstructed well by TOM (see Figure \ref{fig:slf_tractseg}), but semi-automated approaches did not perform as well in this large tract with numerous complex features.

\begin{figure}
    \centering
    \includegraphics[width=1.0\linewidth]{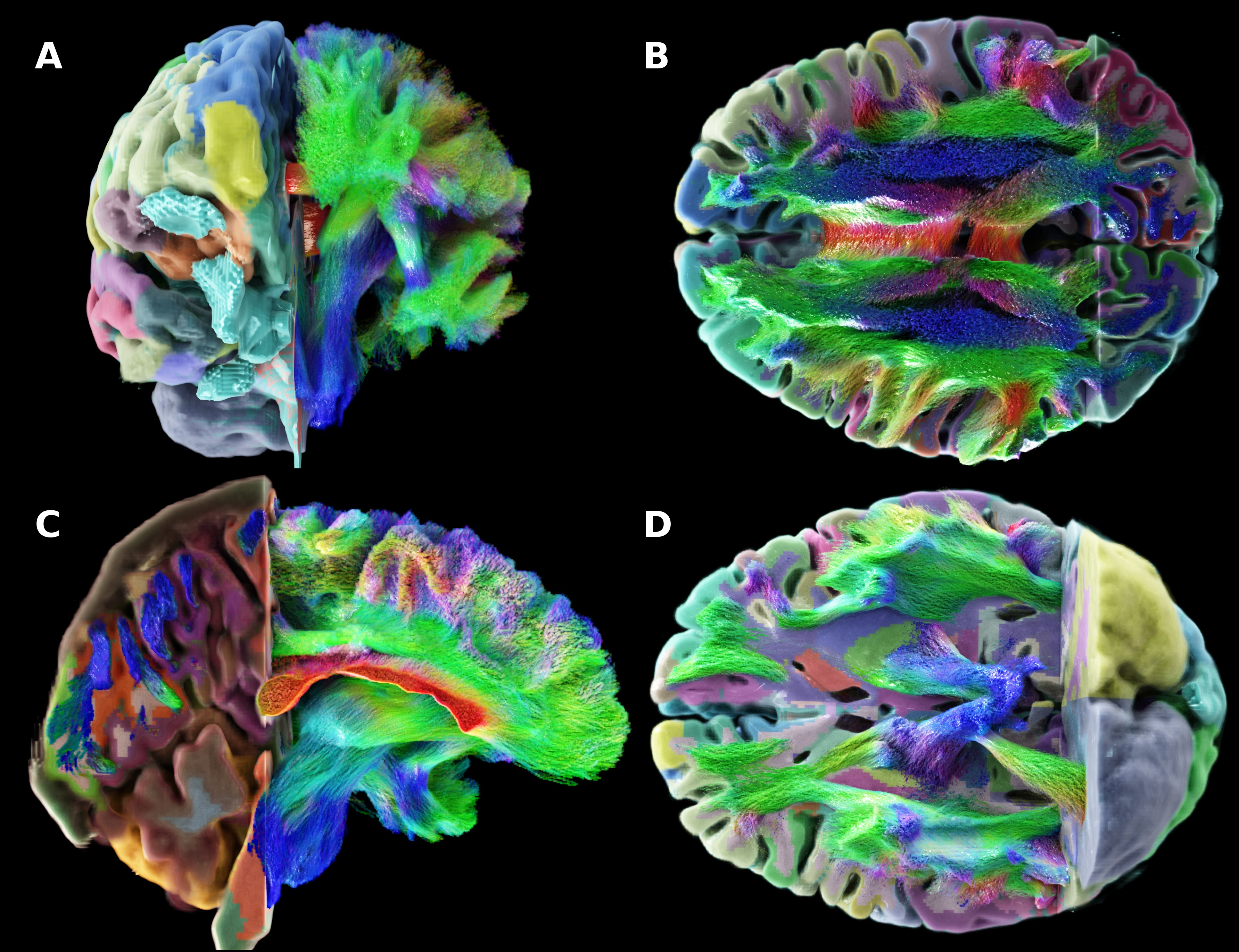}
    \caption{Cinematic renderings of wholebrain tractography obtained using TractSeg, showing how they intersect with the cortical parcellation from FreeSurfer. The combination of tractography and functional parcellation allows estimation of whole-brain connectivity~\cite{Planchuelo-Gomez2024}. Coronal 3D views such as (A) highlight that corticospinal structure is inherently retrieved when using large FOV 3D imaging. (D) shows even cerebellar white matter pathways are mapped despite their relatively low diffusion anisotropy. (B) shows detailed depiction of association, projection and commissural pathways, while (C) clearly demonstrates the separation of the cingulum from the callosum. }
    \label{fig:siemens_visuals}
\end{figure}

\subsection{Artifacts}
Significant artifacts were observed in all ULF scans conducted, and were of varying consequence in subsequent post-processing. Major artifacts and their remediation (if available) were:
\begin{itemize}
    \item Partial voluming arose from the large voxel size employed in this study - this was controlled through the application of the multi-tissue CSD technique, allowing large voxels to remain informative of fibre behaviour.
    \item Diffusion encoding nonuniformity arising from the large  inhomogeneous $B_0$ field gradient relative to the encoding gradients. This was addressed with the magnitude correction approach outlined in Methods and Appendix A.
    \item Broad point spread functions from $T_2$ attenuation during long readouts - these were addressed solely by measures to shorten the readout duration.
    \item Smoothly spatially varying intensity modulation which differed between volumes, possibly due to eddy currents or motion. These were not addressed in the present study.
\end{itemize}

\begin{figure}
    \centering
    \includegraphics[width=1.0\linewidth]{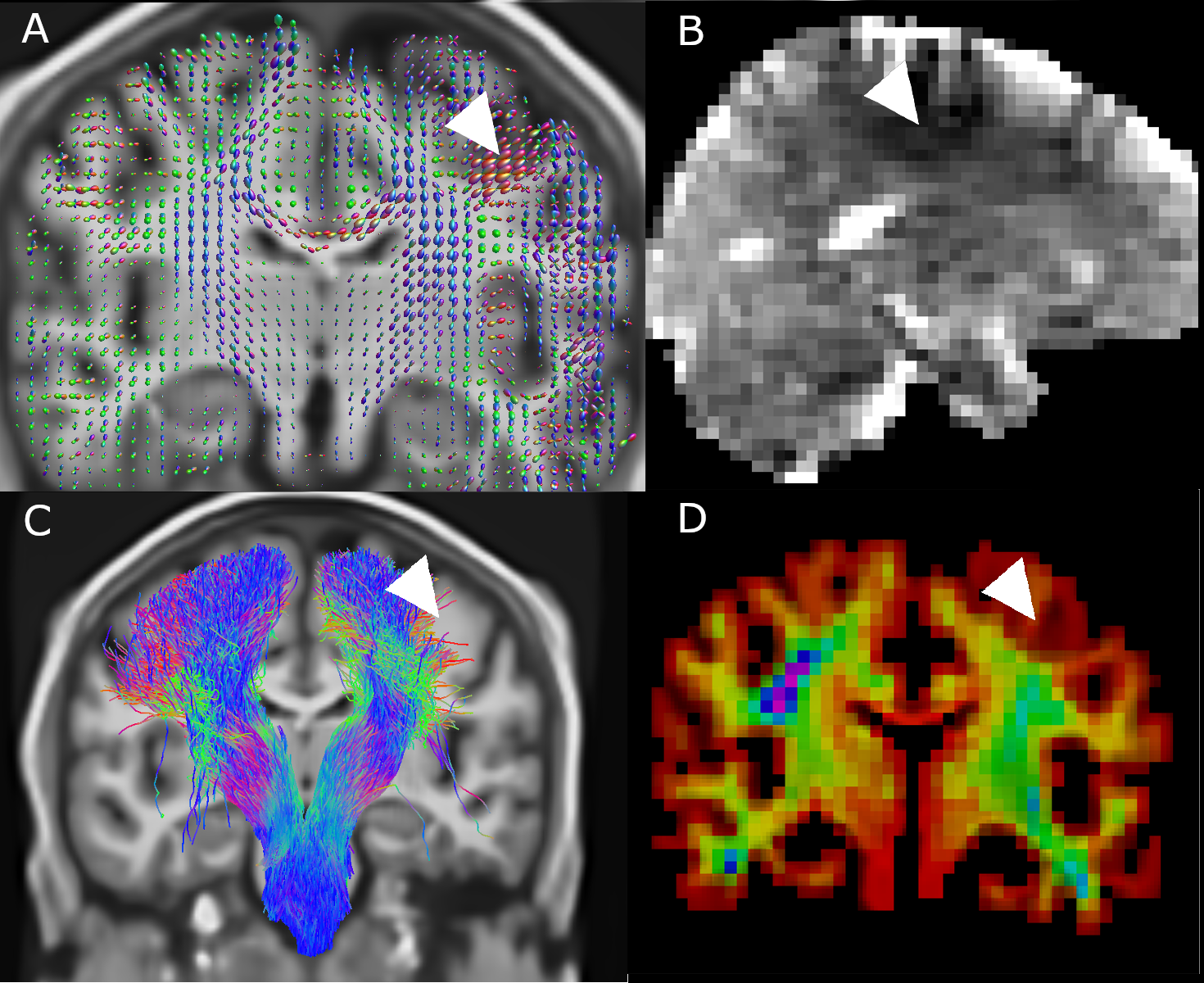}
    \caption{Localised artifact (white arrow) in fODF maps (A) and corresponding hypointensity visible in mean diffusivity images (B). The hypointensity visible in MD maps is diffuse but well localised, suggesting this defect may be readily characterised and corrected. The artifact produced pronounced asymmetry in iFOD/ACT tracking of the CST (C), and corresponding wholebrain track density maps (D) corroborate the defect.}
    \label{fig:hypointensity}
\end{figure}

Partial voluming in the large voxels used is inevitable, an issue exaggerated by movement during the long acquisition. The effect of this is apparent in the reduced fractional anisotropy in narrow fibres measured at ULF vs HF, as well as in defects visible in e.g. the splenium where the narrow tract coincides with the adjacent ventricle (see Figure \ref{fig:hf_lf_quant}). Regions with high CSF content showed apparent fibre content inconsistent with high field reference scans and known anatomy (e.g. a the midsagittal surface as in Figure \ref{fig:hypointensity}A). 

The poor spatial localisation was exacerbated by the broad PSF displayed by FSE sequences with high ETL and low bandwidth, and resulted in FA values near to zero in fibres distant from the medial plane, as well as ambiguous estimates of the principal orientation of the diffusion tensor (as in Figure \ref{fig:hf_lf_quant}A). As well, fODFs in thinner tracts such as the cingulum were visibly diverted by signal contamination from adjacent voxels (Figure \ref{fig:hf_lf}).

Major artifacts were observed in ODF maps in the form of spatially nonuniform responses, and is immediately visible as an asymmetric response between hemispheres (see Figures \ref{fig:hypointensity}A). This varies spatially over relatively large length scales (\SIrange{15}{40}{\milli \meter}). This artifact also occurs in DTI-based diffusion-encoded colour (Figure \ref{fig:hf_lf_quant}C) and mean diffusivity maps (Figures \ref{fig:hf_lf_quant}E and \ref{fig:hypointensity}B) suggesting that nonuniformity in response is not related to multi-tissue CSD. This artifact resulted in a pronounced drop in tract density within the affected region (Figures \ref{fig:hypointensity}, lower left and right).


\section{Discussion}
This work demonstrates that tractography is achievable even on  portable, ULF point-of-care systems. Furthermore, it establishes the viability of 3D diffusion-weighted fast spin echo (DW-FSE) as a tractography sequence at ULF, overcoming significant challenges such as low signal-to-noise ratio (SNR), poor magnet homogeneity, and long, motion-sensitive readouts. The utility of ULF-derived tractography is likely in providing lower-cost, low SAR research imaging at scale, for instance to provide tract specific quantitative measures suitable for measurement at ULF such as magnetisation transfer~\cite{Balaji2025} and quantitative $T_2$ and $T_1$~\cite{Lena2025} metrics, as well as diffusion measures that are available by nature of the experiment.

While we report the best results to date, errors in computed mean diffusivity, fractional anisotropy, and ODF scale and shape parameters highlight that artifacts as well as noise may confound downstream metrics. More artifact-immune sequences employing e.g. more sophisticated navigation or motion correction may reduce these issues to being restricted to noise alone. While the results demonstrated are encouraging, they also highlight the scale of the challenge facing ULF-based tractography. Many tools designed for high-field diffusion pre- and post-processing are employed in this work and found to translate well to ULF (such as registration, bias correction, automated segmentation and multi-tissue CSD), whereas other tools (such as nonlinear registration schemes and Gibbs deringing) were found to work inconsistently.

Several techniques employed in diffusion registration were tested in the development of this pipeline. Affine registration was found to be effective and robust, with common hierarchical approaches functioning as intended.  The low resolution of the base DWIs masked poor registration to a significant degree, and indications of thermal drift producing nonidentical deformations were observed. The low consistency between DWIs from e.g. variation in the presentation of motion or eddy current artifacts likely exacerbated the other issues with registration. Registration of structural images to FA template images was found to produce non-overlapping skull outlines. Nonlinear registration schemes resulted in non-smooth normalisations, particularly in the skull and brain boundaries. Highly regularised nonlinear registration to enforce the constraint of long-range spatially smooth distortions may improve the robustness of these methods.

Signal hyperintensities in the DWI data led to underestimation of mean diffusivity and issues with fODF estimates. The broad spatial extent of these effects points to several potential sources. One candidate is magnetic field nonuniformity—either in the gradient fields, $B_1$, or $B_0$—which typically varies smoothly across the image. Another possibility is the presence of uncorrected gradient moments resulting from eddy currents. These can induce signal loss through residual gradient spoiling, with their impact expected to vary depending on the relative orientation of the diffusion-encoding direction and the readout direction. Notably, the artifact produces directional biases in the diffusion signal that are most pronounced in the lateral directions—coincidentally, also the readout direction—suggesting a systematic interaction between the acquisition scheme and underlying hardware imperfections.

Constrained spherical deconvolution proved highly effective as a model of fibre orientation vs tensor approaches~\cite{Plumley2022}, even when employing relatively sophisticated tensor fitting schemes such as RESTORE~\cite{Chang2005, Chang2012}, and permitted the occurrence of crossing and dispersion fibres, which contributed to the quality of the tractography possible.

Compatibility of the data with automated tractography approaches would be desirable, as it standardises an otherwise subjective workflow, as well as greatly speeding up an otherwise technical and slow process, however the encouraging results shown must be validated against high field data in a range of populations before this approach can be endorsed. Visual inspection suggests a high degree of correspondence with more traditional approaches to tractography like ACT, with the prior information that tracts are large coherent bundles clearly acting to regularise otherwise diffuse tracking. How these approaches may bias results though is as yet untested.

The low inherent SNR and $B_0$ homogeneity of ULF-MRI requires the use of multi-shot imaging methods. DW-FSE provides interpretable, relatively robust diffusion encoded imaging with long echo trains providing many measurements per excitation. $T_2$ attenuation which occurs during long readouts results in significant blurring arising from the reduced echo amplitude at the k-space periphery. This results in a echo-train length (ETL) dependant point-spread function~\cite{Zhou1993}, which, alongside the low spatial resolution, prevents the recovery of smaller tracts or those in dense WM regions.

While diagnostic ULF-DWIs typically use high resolution in-plane and thick slices, diffusion tractography instead benefits from isotropic resolution so as not to produce biased results in certain directions~\cite{Oouchi2007}. Reduction of the in-plane resolution to make voxels isotropic benefits k-space SNR, as this is higher at more central $k$-space locations than more peripheral ones. This facilitates the use of longer echo trains than might otherwise be used for diagnostic sequences, and this work found extended ETLs significantly beyond those typical at ULF conferred substantial benefit to SNR-per-unit-time.

Corrections are demonstrated that address some of the issues that prevent immediate use of data for quantitative DWI, and identify other issues that remain to be addressed. Significant work is still required to address some of the defects observed in the data, particularly spatial variation from gradient nonuniformity and static field gradients. Expectations of encoding uniformity that are typical at high field are shown to be erroneous in DW-ULF, and current tools for signal analysis fail to address this concern. 

In conclusion, this study represents the first successful demonstration of anatomically faithful reconstructions of white matter pathways at ultra-low (64mT) field strength. With the rapid proliferation of low-field MRI systems in low- and middle-income countries (LMICs), this breakthrough has the potential to revolutionize neuroimaging, enabling the study of white matter architecture in regions where it was previously inaccessible. Having established the feasibility of tractography at ultra-low field strengths, future research will focus on enhancing the robustness of microstructural parameter quantification (tractometry) within these tracts. This is where we foresee the greatest impact, as it will enable truly democratized access to advanced neuroimaging techniques, facilitating broad applications in both research and clinical settings.

\section{Acknowledgments}
This work was funded in part by:
\begin{itemize}
    \item The Gates Foundation under the UNITY programme (INV-047888).  
    \item Wellcome LEAP (under the 1kD programme).
    \item A Wellcome Trust Strategic Award (104943/Z/14/Z), 
    \item Wellcome Discovery Awards 227882/Z/23/Z and 317797/Z/24/Z 
\end{itemize}
For the purpose of open access, the author has applied a CC BY public copyright licence to any Author Accepted Manuscript version arising from this submission.\\
\\
Steven Williams would like to thank the National Institute for Health and Care Research (NIHR) Maudsley Biomedical Research Centre (BRC) for ongoing support of our neuroimaging research. \\
\\
 We thank Klaus Engel and Kaloian Petkov of Siemens Healthineers AG, Germany for producing the cinematic renderings in figure \ref{fig:siemens_visuals}.
\clearpage

\section{Appendix}
Presuming the spatially dependent inhomogeneous field $B_{0}(\mathbf{r})$ contributes a significant spatially dependent gradient:

\begin{equation}
    \mathbf{G}(\mathbf{r}) = \frac{\partial B_{0}(\mathbf{r})}{\partial x} + \frac{\partial B_{0}(\mathbf{r})}{\partial y} + \frac{\partial B_{0}(\mathbf{r})}{\partial z}
\end{equation}

we may consider an expression for a perturbation to the effective b-value and direction of the gradient during diffusion encoding. The effective encoding direction is given by

\begin{equation} \label{g_e}
    \mathbf{G}_{e}(\mathbf{r}, t) = \mathbf{G}_{D}(t) +\mathbf{G}(\mathbf{r})
\end{equation}

where $\mathbf{G}_{D}(t)$ is the prescribed, time-varying diffusion gradient and $\mathbf{G}_{e}(\mathbf{r}, t)$ is the effective diffusion gradient. Assuming encoding is dominated by that which occurs while $\mathbf{G}_{D}(t)$ plays, we can readily model the contribution to individual terms of the spatially dependent b-matrix $b_{ij}(\mathbf{r})$. We describe the spatially varying phase encoding given a gradient vector $\mathbf{G}(\mathbf{r}, t)$ as $\mathbf{F}(\mathbf{r}, t)$:

\begin{equation}
    \mathbf{F}(\mathbf{r}, t) = \int_{0}^{TE}\mathbf{G}_{e}(\mathbf{r}, t')dt'
\end{equation}

which is used to compute the pairwise components of the b-matrix in the scanner's frame of reference:

\begin{equation} \label{eq:g2}
    b_{ij}(\mathbf{r}) = \int_{0}^{TE}\mathbf{F}_{i}(\mathbf{r}, t) \mathbf{F}_{j}(\mathbf{r}, t)dt
\end{equation}

We can see from this that the effect of a static field gradient on typical Stejskal-Tanner encoding is to both rotate and scale the encoding gradient. Rotation effects are relatively small for large ratios of $G_D$ to $\mathbf{G}(\mathbf{r})$, even for orthogonal gradients. A large inhomogeneity field with the static gradient orthogonal to the prescribed gradient, with a length ratio of 20:1 would produce a rotation of the effective encoding of \SI{2.9}{\degree}. Comparatively, scaling effects on the interpretation of diffusion encoding are comparatively significant and more pronounced when the prescribed and static gradients are colinear. Using the $\mathbf{G}^2$ scaling implied by equation \ref{eq:g2}, the true encoding magnitude will vary with the quadratic of equation \ref{g_e}. The effective scaling of the encoding magnitude from that prescribed may be considered as a scalar $a$ multiplying the $b$-value, i.e.

\begin{equation}
    b_{\text{effective}} = ab
\end{equation}

And this expression used to modify the Stejskal-Tanner signal equation:

\begin{equation}\label{eq:diff_correction}
    S' = S_{0}\exp({-abD}) \quad \text{hence} \quad S= \exp\left(\log\left(\frac{S'}{S_0}\right) \left(1 - \frac{1}{a}\right)\right)
\end{equation}

i.e. it is possible to modify the measured signal given knowledge of $S_0$ and the scaling factor $a$, which scales approximately with the ratio $(1+\frac{G_{D}}{\mathbf{G}(\mathbf{r})})^{2}$. The actual impact of this effect then is highly dependent on both the diffusion encoding gradient used and the severity of the inhomogeneity.

Though the correction demonstrated is suitable for magnitude correction of encoding errors, we may alternatively solve for a spatially dependent $b$-matrix, as is used for magnetic field gradient inhomogeneity correction. Though this method is more general, it does not readily extend to CSD where a single response function representative of all voxels is required.

\printbibliography
\end{document}